\documentclass[aps,twocolumn,a4paper]{revtex4}
\usepackage{graphicx}
\usepackage{dcolumn}
\usepackage{bm}
\usepackage{amssymb}
\usepackage{amsmath}
\usepackage{epsfig}

\begin{document}

\title{Disembodiment of Physical Properties by Pre- and Post-Selections}
\author{Chang-Ling Zou}
\author{Xu-Bo Zou}
\email{xbz@ustc.edu.cn}
\author{F.-W. Sun}
\email{fwsun@ustc.edu.cn}
\author{Guang-Can Guo}
\affiliation{Key Lab of Quantum Information, University of Science and Technology of
China, Hefei 230026}
\date{\today}

\begin{abstract}
The detailed study of disembodiment of physical properties by pre- and
post-selection is present. A criterion is given to disembody physical
properties for single particle with multiple degrees of freedom. It is shown
that the non-commute operators can also be well separated in different
paths. We generalize the disembodiment to entangled particles, and found
that the disembodiment can happen under special conditions due to the
entanglement.
\end{abstract}

\maketitle

\section{Introduction}

Since the foundation of quantum physics, the controversies about the
self-consistency, completeness, causality and locality in quantum physics
have not stopped. It has shown that the quantum physics is wacky and hard to
interpreted, but have been confirmed by experiments for nearly one century.
Physicists are keeping trying to understand this wacky quantum worlds with a
lot of paradoxes, such as the EPR paradox, Zeno paradox \emph{etc}. These
paradoxes reveal the contradiction of intuition of people and the truth of
nature, and usually inspires the new ideas and new interpretations, further
improve the development of quantum physics.

One of the paradoxes, the pre- and post-selection (PPS) which was firstly
proposed by Aharonov, Bergmann, and Lebowitz (ABL) \cite{abl,PT}, questioned
about the quantum arrow of time. Further studies on PPS leads to numbers of
paradoxes, such as the famous Aharonov, Alber and Vaidman (AAV) paradox
which revealed that the outcome of measurement can be widely out of the
range of the eigenvalues of a system through PPS \cite{aav}. The paradox
about PPS is not only of theoretical interests on foundations of quantum
mechanics, but also offers a powerful tool in experiments. The quantum state
estimation and precision measurement based on the PPS has been proposed, and
successfully demonstrated in experiments \cite{Hosten,Lundeen}. Very
recently, Aharonov, Popescu and Skrzypczyk (APS) proposed that the Cheshire
cats, i.e. the \textquotedblleft body" and \textquotedblleft grin" of cat
can be surprisedly separated through appropriate selections of initial and
final states \cite{Cheshire}. It is very potential for further application
in theoretical and experimental studies \cite{twin}, but the details of the
embodiment of physics properties is still waiting to be explored.

In this paper, we present a general treatment to disembodiment of physical
properties by PPS and give a criterion to disembody physical properties for
single particle with multiple degrees of freedom. It is shown that the
non-commute operators can also be separated successfully. We generalize the
disembodiment to entangled particles, and found that the disembodiment can
happen for special conditions due to the entanglement.

\section{The Cheshire cats and Weak Measurement}

The quantum mechanics is time symmetric, that the initial state is as
important as the final state. Thus, we can prepare particles in selected
initial state $\left\vert \Psi \right\rangle $ which called the
pre-selection, and then postselect the ensemble of the particles
corresponding to a final state $\left\vert \Phi \right\rangle $ through the
measurement in detectors. It has been shown in Ref.\cite{Cheshire}, in the
specific ensembles with PPS, the photon number and the polarization can be
separated in different paths. However, the polarization and photon number
cannot be directly readout through traditional collapse measurement, where
the quantum state is changed significantly. In this case, we cannot
distinguish whether the Cheshire cats is found. If we resorted to the weak
measurement that the ancilla measuring device weakly coupling to the system,
therefore the disturbance to the state of system induced by the measurements
can be neglected. The Hamiltonian of the weak measurement reads

\begin{equation}
H_{I}=\hbar gAO,
\end{equation}%
where $g$ is the interaction strength, $A$ is the ancilla, and $O$ is the
observer operator of the system. In the case of the PPS, the average outcome
(also called Weak Value) of the observer should be

\begin{equation}
\left\langle O\right\rangle _{w}=\frac{\left\langle \Phi |O|\Psi
\right\rangle }{\left\langle \Phi |\Psi \right\rangle },
\end{equation}%
with $\left\langle \Phi |\Psi \right\rangle \neq 0$ that the initial and
final state are not orthogonal to each other. The $\left\langle
O\right\rangle _{w}$ is amplified if the $\left\langle \Phi |\Psi
\right\rangle $ is a very small, which can greatly enhance the measurement
precision but with the scarifies of counts at detector.

Here we want to separate the properties described by operator $O^{j}$ with
superscript $j=1,2,3\ldots $, for different quantum properties.
Disembodiment of a physics properties requires that the expect value of
operator $O^{j}$ is nonzero only in one output, for example, in one output
path for a photon. Additionally, in this path, the expect value of other
operator should be zero. Without loss of generality, to separate operators $%
O^{j}$ in path $j$, the PPS must satisfy

\begin{equation}
\left\langle O_{i}^{j}\right\rangle _{w}=\frac{\left\langle \Phi
|O_{i}^{j}|\Psi \right\rangle }{\left\langle \Phi |\Psi \right\rangle }%
=a_{i}\delta _{ij},
\end{equation}%
where the subscript $i$ denote the path, $a_{i}$ are non-zero numbers and $%
\delta _{ij}$ is Kronecker delta.

\section{Single Particle}

For any single particle system, the quantum states can be present by a
vector in the Hilbert space with the basis $\{\left\vert
b_{i}^{jk}\right\rangle \}$, where $i=1,\ldots ,p$ and $j=1,\ldots ,q$ stand
for different paths and physical properties, and $k=1,2,\ldots ,d$ show the
degrees of freedom of the $j$th physical properties. For path $i$, the PPS
can be present as $\left\langle \Phi _{i}\right\vert
=\{x_{i}^{1},x_{i}^{2},...,x_{i}^{n}\}$ and $\left\vert \Psi
_{i}\right\rangle =\{y_{i}^{1},y_{i}^{2},...,y_{i}^{n}\}^{T}$, where the
dimension $n=q\times d$. The condition for disembodiment in path $i$ becomes

\begin{equation}
\left\langle \Phi _{i}\right\vert O_{i}^{j}\left\vert \Psi _{i}\right\rangle
=a_{i}\delta _{ij},
\end{equation}%
with the weak value unnormalized. Or we can write the equation in the matrix
form as

\begin{equation}
\sum_{kl}(O_{i}^{j})_{kl}x_{i}^{k}y_{i}^{l}=a_{i}\delta _{ij},
\end{equation}%
where $(O_{i}^{j})_{kl}=\left\langle b_{i}^{jk}\right\vert
O_{i}^{j}\left\vert b_{i}^{jl}\right\rangle $. Combining all operators $%
j=1,\ldots ,q$, we can write the tensor in the matrix form as

\begin{eqnarray}
&&M_{i}\{x_{i}^{1}y_{i}^{1},x_{i}^{1}y_{i}^{2},\ldots ,x_{i}^{n}y_{i}^{n}\}
\\
&=&\{0,\ldots ,0,a_{i},0,\ldots ,0\}^{T},
\end{eqnarray}%
where the dimension of $M_{i}$ is $q\times n^{2}$. In the present case, the
physical properties and operators are the same for all paths, thus we have $%
M_{i}=M$ independent on the path. Suppose there are $m$ operators to
separate ($m\leq p$), then we should solve the linearized equations

\begin{eqnarray}
M\overrightarrow{v_{1}} &=&\{a_{1},0,\ldots ,0\}^{T},  \notag \\
M\overrightarrow{v_{2}} &=&\{0,a_{2},0,\ldots ,0\}^{T}\text{.} \\
&&\ldots   \notag
\end{eqnarray}

The criterion of the existance of solutions to above equations is

\begin{equation*}
\mathrm{rank}(M)=m.
\end{equation*}

Applying this criteria, we study two examples of photon with polarization
degree of freedom$_{{}}$.

\textbf{Example I}: Single photon with the polarization $\left\vert
+\right\rangle $ or $\left\vert -\right\rangle $ in two paths $\left\vert
1\right\rangle $ or $\left\vert 2\right\rangle $. The operator of the which
photon number operator is

\begin{equation}
O_{i}^{1}=I_{i}=diag\{1,1\},
\end{equation}%
and the polarization operator is

\begin{equation}
O_{i}^{2}=\sigma _{i}^{z}=diag\{1,-1\}.
\end{equation}

Then, we have the matrix $M=\left(
\begin{array}{cc}
1 & 1 \\
1 & -1%
\end{array}%
\right) $, with $\mathrm{rank}(M)=2$. Thus, we can separate the photon
number and polarization in two different paths. We need to solve the
equations

\begin{eqnarray}
M\left(
\begin{array}{c}
x_{1}^{1}y_{1}^{1} \\
x_{1}^{2}y_{1}^{2}%
\end{array}%
\right)  &=&\left(
\begin{array}{c}
1 \\
0%
\end{array}%
\right) \text{,} \\
M\left(
\begin{array}{c}
x_{2}^{1}y_{2}^{1} \\
x_{2}^{2}y_{2}^{2}%
\end{array}%
\right)  &=&\left(
\begin{array}{c}
0 \\
1%
\end{array}%
\right) \text{,}
\end{eqnarray}%
and we have $x_{1}^{1}y_{1}^{1}=x_{1}^{2}y_{1}^{2}=\frac{1}{2}$ and $%
x_{2}^{1}y_{2}^{1}=-x_{2}^{2}y_{2}^{2}=\frac{1}{2}$. There are infinite
choices of pre-selections and post-selections. Let the pre-selection state
be $\left\langle \Psi \right\vert
=\{x_{1}^{1},x_{1}^{2},x_{2}^{1},x_{2}^{2}\}=\{\frac{1}{2},\frac{1}{2},\frac{%
1}{2},\frac{1}{2}\}$, then $\left\vert \Phi \right\rangle
=\{y_{1}^{1},y_{1}^{2},y_{2}^{1},y_{2}^{2}\}^{T}=\{\frac{1}{2},\frac{1}{2},%
\frac{1}{2},-\frac{1}{2}\}^{T}$. This is exactly the case in the original
Cheshire Cats paper by Aharonov et al. \cite{Cheshire}.

\textbf{Example II}: Single photon with polarization $\left\vert
+\right\rangle $ or $\left\vert -\right\rangle $ in four path $\left\vert
1\right\rangle ,\left\vert 2\right\rangle ,\left\vert 3\right\rangle $ or $%
\left\vert 4\right\rangle $. We want to separate the photon number operator $%
I$ and the polarization operators $\sigma ^{x},\sigma ^{y}$ and $\sigma ^{z}$
in different paths. Thus, we have

\begin{eqnarray}
I_{i} &=&\left(
\begin{array}{cc}
1 & 0 \\
0 & 1%
\end{array}%
\right) , \\
\sigma ^{x} &=&\left(
\begin{array}{cc}
0 & 1 \\
1 & 0%
\end{array}%
\right) , \\
\sigma ^{y} &=&\left(
\begin{array}{cc}
0 & -i \\
i & 0%
\end{array}%
\right) , \\
\sigma ^{z} &=&\left(
\begin{array}{cc}
1 & 0 \\
0 & -1%
\end{array}%
\right) ,
\end{eqnarray}

The matrix form is

\begin{equation}
M=\left(
\begin{array}{cccc}
1 & 0 & 0 & 1 \\
0 & 1 & 1 & 0 \\
0 & -i & i & 0 \\
1 & 0 & 0 & -1%
\end{array}%
\right) \text{.}
\end{equation}

In path $j$, the equation should be satisfied

\begin{equation}
M\left(
\begin{array}{c}
x_{j}^{1}y_{j}^{1} \\
x_{j}^{1}y_{j}^{2} \\
x_{j}^{2}y_{j}^{1} \\
x_{j}^{2}y_{j}^{2}%
\end{array}%
\right) =\left(
\begin{array}{c}
\delta _{1j} \\
\delta _{2j} \\
\delta _{3j} \\
\delta _{4j}%
\end{array}%
\right) .
\end{equation}

Since $\mathrm{\det }(M)=-4i$, thus $\mathrm{rank}(M)=4$, we can separate
all four operators in four paths. One set of the solutions is

\begin{eqnarray}
x_{1}^{1}y_{1}^{1} &=&x_{1}^{2}y_{1}^{2}=\frac{1}{2},  \notag \\
x_{2}^{1}y_{2}^{2} &=&x_{2}^{2}y_{2}^{1}=\frac{1}{2},  \notag \\
x_{3}^{1}y_{3}^{2} &=&-x_{3}^{2}y_{3}^{1}=\frac{i}{2},  \notag \\
x_{4}^{1}y_{4}^{1} &=&-x_{4}^{2}y_{4}^{2}=\frac{1}{2}.
\end{eqnarray}

Thus, we can have $\left\langle \Psi \right\vert =\{1,1,1,1,1,1,1,1\}$ and $%
\left\vert \Phi \right\rangle =\{1,1,1,1,i,-i,1,-1\}^{T}.$

\section{Entangled Particles}

Now, we turn to consider the two-particle system. If there is no
entanglement, the two-particle system is only the simple extension of single
particle systems with more degrees of freedom. Take the two photons in four
path (photon A in path $\left\vert 1\right\rangle $ or $\left\vert
2\right\rangle $; photon B in path $\left\vert 3\right\rangle $ or $%
\left\vert 4\right\rangle $ ) with the the polarization is entangled ($%
\left\vert H\right\rangle _{A}\left\vert V\right\rangle _{B}+\left\vert
H\right\rangle _{B}\left\vert V\right\rangle _{A}$). Thus, the Hilbert space
of state can be represent in the basis

\begin{eqnarray}
&&\{\left\vert 1H\right\rangle \left\vert 3V\right\rangle ,\left\vert
1H\right\rangle \left\vert 4V\right\rangle ,\left\vert 1V\right\rangle
\left\vert 3H\right\rangle ,\left\vert 1V\right\rangle \left\vert
4H\right\rangle ,  \notag \\
&&\left\vert 2H\right\rangle \left\vert 3V\right\rangle ,\left\vert
2H\right\rangle \left\vert 4V\right\rangle ,\left\vert 2V\right\rangle
\left\vert 3H\right\rangle ,\left\vert 2V\right\rangle \left\vert
4H\right\rangle .\}
\end{eqnarray}

Since the polarization is entangled, the dimensional of the state space is
8. The operators of the photon number are

\begin{eqnarray}
I_{1}^{A} &=&diag\{1,1,1,1,0,0,0,0\},  \notag \\
I_{2}^{A} &=&diag\{0,0,0,0,1,1,1,1\},  \notag \\
I_{3}^{B} &=&diag\{1,0,1,0,1,0,1,0\},  \notag \\
I_{4}^{B} &=&diag\{0,1,0,1,0,1,0,1\}.
\end{eqnarray}

and operators of photon polarization are

\begin{eqnarray}
\sigma _{1}^{A} &=&diag\{1,1,-1,-1,0,0,0,0\},  \notag \\
\sigma _{2}^{A} &=&diag\{0,0,0,0,1,1,-1,-1\},  \notag \\
\sigma _{3}^{B} &=&diag\{-1,0,1,0,-1,0,1,0\},  \notag \\
\sigma _{4}^{B} &=&diag\{0,-1,0,1,0,-1,0,1\}.
\end{eqnarray}

The matrix form is

\begin{equation}
M=\left(
\begin{array}{cccccccc}
1 & 1 & 1 & 1 & 0 & 0 & 0 & 0 \\
0 & 0 & 0 & 0 & 1 & 1 & 1 & 1 \\
1 & 0 & 1 & 0 & 1 & 0 & 1 & 0 \\
0 & 1 & 0 & 1 & 0 & 1 & 0 & 1 \\
1 & 1 & -1 & -1 & 0 & 0 & 0 & 0 \\
0 & 0 & 0 & 0 & 1 & 1 & -1 & -1 \\
-1 & 0 & 1 & 0 & -1 & 0 & 1 & 0 \\
0 & -1 & 0 & 1 & 0 & -1 & 0 & 1%
\end{array}%
\right) .
\end{equation}

We find that, for the expecting value for operator as

\begin{equation}
\overrightarrow{e}=\{1,0,0,1,1,0,0,1\}^{T},
\end{equation}%
the solution to

\begin{equation}
M\overrightarrow{v}=\overrightarrow{e}
\end{equation}%
does not exist. However, for

\begin{equation}
\overrightarrow{e}=\{1,0,0,1,1,0,0,-1\}^{T},
\end{equation}%
the disembodiment can happen. This is due to the polarization entanglement
between the two photons, where the expecting values should be $1$ and $-1$
respectively. So, we can separate the physics properties even for entangled
particles.

\section{Discussion}

(1) From the analysis for single particle above, the disembodiment of
physical properties is not restricted to the system with two degrees of
freedom. The physics properties can always be separated through particular
PPS ensemble according to the criterion. In addition, the selected initial
and final states for disembodiment are not sole.

(2) The disembodiment is not restricted to separation physical properties in
different paths. It can be extended to any other degree of freedom that we
can address in experiment, such as internal degree of atoms.

(3) Potential application of disembodiment would be selectively measure the
parameters with different operators. For instance, for a system interact
with ancilla $H_{I}=\hbar A(g_{1}O^{1}+g_{2}O^{2})$, where $O^{1}$ and $%
O^{2} $ are independent to each other, the observer $O^{1}$ and $O^{2}$ can
be selectively measured through disembodiment by only one ancilla.

(4) In all above analysis, the PPS is perfect regardless the actual
preparation and detection process. The real experimental situation, the
imperfect devices give errors in state preparation and detection. This type
of error is stationary and can be modified through transformations when the
imperfection of devices are calibrated \cite{shen}. Noting that, the effect
of noise is random that can not be estimated through transformation.

\section{Conclusions}

In summary, we have studied the disembodiment of physical properties by pre-
and post-selection in detail. We give a criterion to disembody physical
properties for single particle with multiple degrees of freedom. It is shown
that the non-commute operator can also be separated in different paths. We
generalize the disembodiment to entangled particles and find that the
disembodiment can happen for special conditions due to the entanglement.

\textbf{Acknowledgements} This work was supported by the 973 Program under
Grant 2011CB921200 and Grant 2011CBA00200, the Natural Science Foundation of
China under Grant 11004184, and in part by the Knowledge Innovation Project
of the Chinese Academy of Science.


\begin{thebibliography}{9}
\bibitem{abl} Y. Aharonov, P. G. Bergmann, and J. Lebowitz, Phys. Rev. 134,
B1410 1964 .

\bibitem{PT} Y. Aharonov, S. Popescu, and J. Tollaksen, Phys. Today \textbf{%
63}(11), 27 (2010).

\bibitem{aav} Y. Aharonov, D. Z. Albert, and L. Vaidman, Phys. Rev. Lett.
\textbf{60}, 1351 (1988).

\bibitem{Hosten} O. Hosten and P. Kwiat, Science \textbf{319}, 787 (2008).

\bibitem{Lundeen} J. S. Lundeen, B. Sutherland, A. Patel, C. Stewart, and C.
Bamber, Nature \textbf{474}, 188 (2011)

\bibitem{Cheshire} Y. Aharonov, S. Popescu, and P. Skrzypczyk, arXiv:
1202.0631 (2012).

\bibitem{twin} I. Ibnouhsein, and A. Grinbaum, arXiv: 1202.4894 (2012).

\bibitem{shen} C. Shen and L.-M. Duan, arXiv:1201.4379 (2012).
\end{thebibliography}
\end{document}